\documentclass[12pt]{article}
\usepackage{epsfig}

\makeatletter

 \@addtoreset{equation}{section}
\makeatother

\newcommand {\beq}{\begin{eqnarray}}
\newcommand {\eeq}{\end{eqnarray}}
\newcommand {\non}{\nonumber\\}

\newcommand {\gto}{\stackrel{g}{\to}}

\newcommand {\1}[1]{\frac{1}{#1}}

\newcommand {\ph}{\varphi}

\newcommand {\sig}{\sigma}

\newcommand {\del}{\partial}
\newcommand {\dagg}{^{\dagger}}

\newcommand {\lam}{\lambda}

\newcommand {\GC}{G^{\bf C}}

\newcommand{\vs}[1]{\vspace{#1 mm}}
\newcommand{\hs}[1]{\hspace{#1 mm}}

\newcommand{\kahler}{K\"ahler }

\newcommand{\ib}{\bar{i}}
\newcommand{\jb}{\bar{j}}

\begin{document}

\setcounter{page}{0}

\begin{titlepage}

\begin{flushright}
~\\
{\tt hep-th/0309004}\\
September 2003
\end{flushright}
\bigskip

\begin{center}
{\LARGE\bf
Non-compact Calabi-Yau Metrics \\ 
from Nonlinear Realizations
}
\vs{10}

\bigskip
{\renewcommand{\thefootnote}{\fnsymbol{footnote}}
\large\bf Muneto Nitta\footnote{
E-mail: nitta@physics.purdue.edu}
}

\setcounter{footnote}{0}
\bigskip

{\it
Department of Physics, Purdue University, 
West Lafayette, IN 47907-1396, USA
}

\end{center}
\bigskip

\begin{abstract}
We give a method to construct 
Calabi-Yau metrics on 
$G$-invariant vector bundles 
over K\"ahler coset spaces $G/H$ 
using supersymmetric nonlinear realizations 
with matter coupling.
As a concrete example we discuss 
the ${\bf C}P^N$ model coupled with matter.
The canonical line bundle is reproduced by 
the singlet matter and the cotangent bundle 
with a new non-compact Calabi-Yau metric 
which is not hyper-K\"ahler 
is obtained by 
the anti-fundamental matter.

\end{abstract}

\end{titlepage}

\section{Introduction}

The Einstein equation is a partial differential 
equation and so is very difficult to solve 
in general without symmetry.
When the manifold has isometry large enough, 
it is often reduced to an algebraic equation or 
an ordinary differential equation (ODE) which is easy to solve.
The Einstein manifolds satisfying 
$R_{\mu\nu} = h g_{\mu\nu}$~\cite{Be} 
are solutions of the Einstein equation.
For instance, homogeneous spaces (coset spaces) 
$G/H$ admit Einstein metrics with positive $h$, 
in which case the Einstein equation reduces to 
a set of algebraic equations.

The Calabi-Yau manifolds 
which are important ingredient for 
string compactification~\cite{CHSW}  
are K\"ahler-Einstein manifolds with vanishing $h$ in 
one of their definitions.
They cannot be homogenous because homogeneous manifolds 
have positive $h$, 
so the next class expected to be solved is cohomogeneity one~\cite{DW}
although the manifold is {\it non-compact}.  
Therefore the classification of Calabi-Yau manifolds 
of cohomogeneity one is important and interesting. 
A class is given by the Stenzel metrics on 
the cotangent bundles $T^*(G/H)$ 
with $G/H$ rank one coset spaces~\cite{St}. 
In particular, 
in the case of $G/H = SO(N)/SO(N-1) \simeq S^{N-1}$,  
it is the higher-dimensional 
deformed conifold~\cite{St,CGLP,HKN1}, 
which includes 
the (six-dimensional) deformed conifold~\cite{conifold,conifold2} 
and the Eguchi-Hanson manifold~\cite{EH} 
as lower dimensional manifolds.
Another class is given by
the canonical line bundles over K\"ahler coset spaces 
$G/H$~\cite{Ca,PP}.
For each of these models the Einstein equation 
has enough symmetry to be reduced to ODE.

A natural metric on a conifold 
different from \cite{conifold} 
was constructed and identified 
with the canonical line bundle over the quadric surface 
$Q^N = SO(N+2)/[SO(N) \times U(1)]$~\cite{HKN2,PV}. 
In \cite{HKN3} this was generalized to 
Hermitian symmetric spaces (HSS) $G/H$ 
with classical groups $G$, 
using supersymmetric gauge theories. 
Generalized conifold with $E_6$ ($E_7$) symmetry, 
which is defined by
$\Gamma_{ijk} \phi^j \phi^k = 0$ 
($d_{\alpha\beta\gamma\delta} \phi^{\beta} 
\phi^{\gamma} \phi^{\delta} = 0$) 
with $\Gamma$ ($d$) 
the rank 3 (rank 4) symmetric tensor of 
$E_6$ ($E_7$) and with 
$\phi^i$ ($\phi^{\alpha}$) 
the fundamental representation ${\bf 27}$ (${\bf 56}$), 
was constructed in \cite{HKN4}.
It was identified with 
the canonical line bundle over the exceptional HSS 
$E_6/[SO(10) \times U(1)]$ ($E_7/[E_6 \times U(1)]$). 
Later in \cite{HKN5} we directly 
constructed the canonical line bundle 
over arbitrary K\"ahler-Einstein coset space $M = G/H$ 
using the nonlinear realization method,  
and manifolds obtained in Refs.~\cite{HKN2}--\cite{HKN4} 
were correctly reproduced for HSS.

In this paper, 
we generalize this class to 
$G$-invariant vector bundles over 
K\"ahler coset spaces $M=G/H$, 
using the nonlinear realization method with 
matter coupling~\cite{CCWZ}.
From the requirement of $G$-invariance, 
the dimensions of 
the vectors as fibers are restricted 
because 
they are matter fields 
belonging to a representation of $H$. 
The number of $G$-invariants composed of fields 
coincides with the cohomogeneity of the manifold. 
(It is also related with the number of 
the so-called quasi-Nambu-Goldstone bosons~\cite{Ni}.)
When the matter belongs to 
an irreducible representation of $H$, 
we have one $G$-invariant and 
the total manifold can be cohomogeneity one, 
so the Ricci-flat condition 
is reduced to ODE to be solved. 
As a concrete example, 
we work out the projective space ${\bf C}P^N$.
The singlet matter reduces to 
the line bundle considered previously, 
and matter in the anti-fundamental representation 
provides the cotangent bundle $T^*{\bf C}P^N$ 
which is {\it not} hyper-K\"ahler.
Our model gives new finite nonlinear sigma model and 
therefore suggests a conformal field theory.

This paper is organized as follows.
In Sec.~2, a brief review on 
supersymmetric nonlinear realizations 
with matter coupling is given 
with emphasis on its relation with 
cohomogeneity of the manifold. 
In Sec.~3, we apply this method to 
the ${\bf C}P^N$ model coupled with matter 
and briefly discuss the singlet matter.
A new metric on the cotangent bundle is 
discussed in Sec.~4. 
Sec.~5 is devoted to summary and discussions. 
In Appendix A, we give a relation 
with the hyper-K\"ahler Calabi metric 
on the cotangent bundle 
using the (hyper-)K\"ahler quotient.

\section{Nonlinear Realizations with Matter}
In this section, we review the nonlinear realization 
with matter coupling focusing on 
its relation with our problem.
It provides an easy procedure to construct 
the $G$-invariant metric on $G/H$ as 
a nonlinear sigma model~\cite{CCWZ}.
In supersymmetric nonlinear sigma models 
the associated manifolds must be K\"ahler~\cite{Zu}, 
and we need K\"ahler potentials 
$K(\ph,\ph^*)$ instead of the metrics 
for manifestly supersymmetric Lagrangian:~\footnote{
We use the language of 
the four-dimensional ${\cal N}=1$ supersymmetry 
for superfields~\cite{WB}.
}
\beq
 {\cal L} = \int d^4 \theta \, K(\ph,\ph\dagg) 
 = - g_{ij^*}(\ph,\ph^*) \del_{\mu}\ph^i \del^{\mu}\ph^{*j} 
 + \cdots  
\eeq
with $g_{ij^*} \equiv \del_{i}\del_{j^*} K(\ph,\ph^*)$ 
the K\"ahler metric, 
where dots denote terms including 
fermionic superpartners of $\ph^i$. 
Here we have used the same letter $\ph^i$ 
for chiral superfields and 
their complex scalar components. 
The most general discussion for 
nonlinear realization on target \kahler manifolds 
was given in \cite{BKMU} (see \cite{Ku} as a review).  
They gave systematic method to construct 
the K\"ahler potential invariant under $g \in G$ 
up to a K\"ahler transformation:
$K(\ph ,\ph^*) \gto K(\ph' ,\ph^*{}') 
= K(\ph ,\ph^*) + \Lambda (g,\ph) + \Lambda^*(g,\ph^*)$.
The target manifolds can be 
compact homogeneous \cite{IKK}--\cite{HKNT} or 
non-compact non-homogeneous 
\cite{non-compact,Ni}. 
The matter coupling was discussed in \cite{BKMU,matter,line}.

We would like to construct manifolds 
of cohomogeneity one  
by adding matter to base manifolds $M$. 
Hence we consider 
compact homogenous K\"ahler coset spaces $M = G/H$
as base manifolds because 
non-compact manifolds constructed by nonlinear realizations 
are non-homogeneous so cohomogeneity greater than one.
Compact K\"ahler coset spaces $G/H$ can be written as
$G/H = G/[H_{\bf s.s.} \times U(1)^r]$ 
with $H_{\rm s.s.}$ the semi-simple subgroup in $H$ and 
$r \equiv {\rm rank}\, G - {\rm rank }\, H_{\rm s.s.}$~\cite{Bo}. 
There exists isomorphism 
$G/H \simeq G^{\bf C}/\hat H$ 
where $G^{\bf C}$ is the complexification of $G$ and 
$\hat H$ is its complex subgroup 
of which $H^{\bf C}$ is a subgroup: 
$\hat {\cal H} = {\cal H}^{\bf C} \oplus B$ 
with $B$ nilpotent generators. 
(We use the Calligraphic font for Lie algebras.)
To construct supersymmetric Lagrangian, 
we use the complex coset spaces 
$G^{\bf C}/\hat H$ 
because they are directly parameterized by 
complex coordinates as chiral superfields. 
The coset representative of $\GC/\hat H$ is given by 
$\xi = \exp (\ph \cdot Z)$ 
with $\ph^i$ chiral superfields and 
$Z_i$ complex generators in ${\cal G}^{\bf C} - \hat {\cal H}$. 
It is transformed under $g \in G$ as 
\beq
 \xi \gto 
 \xi' \equiv \exp (\ph' \cdot Z) 
      = g \xi \hat h'{}^{-1}(g,\xi) \,, 
  \label{xi-tr}
\eeq
where $\hat h' \in {\hat H}$ is a compensator 
needed to put $g \xi$ into an element 
of a coset representative as Figure 1.
\begin{figure}[h]
\begin{center}
\leavevmode
  \epsfysize=6cm
  \epsfbox{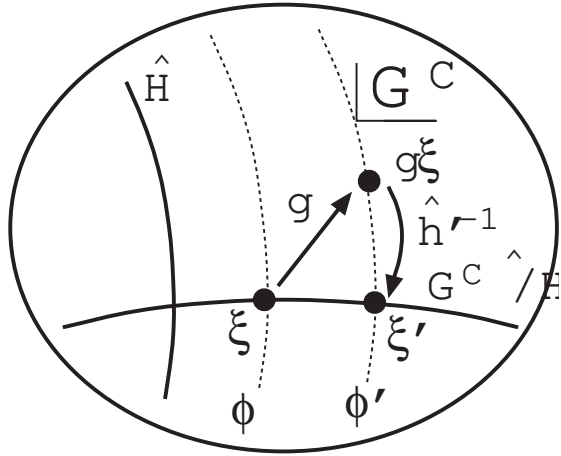} \\ 
\caption{
The $G$-transformation law for $\xi$}
\label{fig1}
\end{center}
\begin{small}

\end{small}
\end{figure}

The K\"ahler potential for K\"ahler $G/H$ invariant under $G$ 
up to a K\"ahler transformation
is given by~\cite{BKMU,IKK,BKY}
\beq
 K = \sum_a c_a \log \det{}_{\eta_a} \xi\dagg\xi \;,
 \label{kahler-coset}
\eeq
with $a=1,\cdots,r$ and $c_a$ real constants. 
We have used the fundamental representation for $\xi$.
$\eta_a$ are projection operators 
satisfying $\eta \hat H \eta = \hat H \eta$, $\eta^2 = \eta$
and $\eta\dagg = \eta$, and
$\det{}_{\eta}$ denotes the determinant 
in the subspace projected by $\eta$.
For ${\bf C}P^N$ discussed in this paper in detail, 
the K\"ahler potential is just 
the one for the Fubini-Study metric:
$K = c \log (1 +|\ph|^2)$.

\medskip

In the nonlinear realization method,
the matter fields are defined as fields belonging to 
some representation of $H$ and 
their nonlinear $G$-transformation can 
be defined in the standard way~\cite{CCWZ}. 
They are usually considered as fermions 
in non-supersymmetric theory.
In supersymmetric theories, matter also belong to 
chiral superfields 
so their bosonic degrees of freedom add 
non-compact directions to $M$. 
The total space becomes a $G$-invariant vector 
bundle over $M$ 
with bosonic matter as a fiber.

Let $\psi$ be matter chiral superfields 
belonging to the representation $\rho_0$ of $H$. 
Now we assume that this representation is {\it irreducible}. 
The nonlinear $G$-transformation of the matter fields 
is defined by 
\beq
 \psi \gto \psi' = \rho_0(\hat  h'(g,\xi)) \psi \,
\eeq
using the compensator $\hat h'$ in (\ref{xi-tr}).
The matter representation $(\psi,\rho_0)$ can be embedded into 
some representation $(\tilde\psi,\rho)$ of $\GC$ 
with simply adding zero components, like 
\beq
 \tilde\psi = \pmatrix{\psi \cr {\bf 0}} \, ,\hs{5} 
 \rho (\hat H) 
 = \pmatrix{\rho_0(\hat H) & \cr
                           & \ddots } \; . 
\eeq
By a field redefinition 
using the coset representative $\xi$, 
matter chiral superfields $\chi$ defined by
\beq
 \chi = \rho(\xi) \tilde\psi \, 
\eeq
is found to transform {\it linearly} under $G$:
\beq
 \chi \gto \chi' = \rho(\xi') {\tilde\psi}' 
 = \rho (g \xi \hat h'{}^{-1}) \rho(\hat h') \tilde\psi 
 = \rho(g) \chi \, \label{matter-tr.} \;,
\eeq
which are called 
the standard representation~\cite{CCWZ}. 

(One of) the $G$-invariant of matter fields can be found immediately
from (\ref{matter-tr.}) as
\beq
 X \equiv |\chi|^2 \equiv \chi\dagg\chi 
 = {\tilde\psi}\dagg\rho(\xi\dagg\xi)\tilde\psi 
 = (\psi\dagg,{\bf 0}) \rho(\xi\dagg\xi) 
   \pmatrix{\psi \cr {\bf 0}} \,.  \label{invariant}
\eeq
The \kahler potential for the matter fields 
can be written as~\footnote{
When there exists a $G$-symmetric tensor 
for this representation, 
we have more $G$-invariants. 
For instance, $G=SO(N)$ has the invariant tensor 
$\delta_{ij}$ so real and imaginary parts of 
$\sum_i \chi^i\chi^i$ are also invariant.
Such a case does not give us a manifold of cohomogeneity one.
} 
\beq
 K_{\rm matter} = f(X) \,,
\eeq
where $f$ is an arbitrary function.  
The K\"ahler potential for 
the total space coupled with the matter fields 
can be written as
\beq
 K = \sum_a c_a \log \det{}_{\eta_a} \xi\dagg\xi + f(X) \,. 
\eeq
If we set $\psi=0$, we have $X= 0$ and 
the \kahler potential reduces to 
the one of the \kahler coset $G/H$. 
Therefore the total space is the vector bundle over 
$G/H$ with the matter $\psi$ fiber.

\medskip
Before closing this section we make a comment on 
the importance of the irreducibility.
If $\rho_0$ is {\it reducible}, 
it can be divided into $n$-irreducible sectors $\rho_0^{(I)}$ of $H$ 
with $I(=1,\cdots,n)$ labeling each $H$-sector.
In the same way, we can construct the matter in 
the standard representation $\chi^{(I)}$. 
There exist at least $n$ $G$-invariants 
$X^{(I)} \equiv |\chi^{(I)}|^2$.
If two of $\chi^{I}$'s are in the same representation 
the inner product of them, 
$Y^{(IJ)} \equiv \chi^{(I)}{}\dagg \chi^{(J)}$, 
is also invariant. 
The matter \kahler potential is an arbitrary function 
of the several variables:
\beq
 K_{\rm matter} = f(X^{(1)}, \cdots, X^{(n)}; 
 Y^{(IJ)} + Y^{(IJ)},  Y^{(IJ)} + i Y^{(IJ)}, \cdots ) \,.
\eeq
Since the number of $G$-invariants coincides with 
the cohomogeneity of the manifold,
this case obviously gives us a manifold 
with cohomogeneity greater than $n$. 
Therefore the Ricci-flat condition is still 
a partial differential equation 
although the number of variables is reduced.

\section{Projective space ${\bf C}P^{N}$ coupled with matter}
We discuss 
${\bf C}P^{N} = SU(N+1)/[SU(N) \times U(1)] = G/H$,
which is parameterized by the fields $\ph^i$ 
belonging to 
the fundamental representation ${\bf N}$ of $SU(N)$. 
For the ${\bf C}P^N$ model, 
the complex isotropy is given by 
$\hat {\cal H} 
= \pmatrix 
 {{\cal U}(1)^{\bf C} &  {\rm B}\;\; \cdots \;\;{\rm B} \cr 
              {\bf 0} & {\cal SU}(N)^{\bf C}}$ 
with B denoting $N$ nilpotent generators. 
The coset representative of ${\bf C}P^{N}$ is given by 
\beq
 \xi = e^{\ph \cdot Z} 
 =  \pmatrix {1  & {\bf 0} \cr 
             \ph & {\bf 1}_{N} } \,, \hs{5} 
 \xi^{-1 T} 
 =  \pmatrix { 1  &  - \ph^T \cr 
          {\bf 0} & {\bf 1}_{N}  } \,. 
\eeq
As the matter fields coupled with ${\bf C}P^N$, 
we can consider 
the singlet ${\bf 1}$, 
the fundamental representation ${\bf N}$ or 
the anti-fundamental representation $\overline{{\bf N}}$ 
of $SU(N)$ with suitable charges for $U(1) \subset H$ 
immediately.\footnote{
Of course, we can consider higher representations 
of $SU(N)$.
}
The total space is the vector bundle over 
${\bf C}P^{N}$ with the matter fields as a fiber. 
The \kahler potential for the total space can be written as 
(the projection $\eta$ 
in Eq.~(\ref{kahler-coset}) is given by 
$\eta = {\rm diag.} (1,0,\cdots,0)$.)
\beq
 K = c \log (1 + |\ph|^2) + f(X) \,, \label{CPN-matter}
\eeq
where the $G$-invariant $X$ is constructed as (\ref{invariant})
using the matter fields  
belonging to some representation of $H$. 
Here we summarize the matter coupling belonging to 
the following representations of $SU(N)$, 
1) $\sig \in {\bf 1}$,
2) $\psi \in {\bf N}$ and 
3) $\psi \in \overline{{\bf N}}$:

\begin{enumerate}

\item 
$\sig \in {\bf 1}$.\\ 
In this case, the matter field $\sig$ can be 
promoted to the standard representation as 
\beq
 && \chi = \xi \pmatrix{ \sig \cr {\bf 0} } 
         = \sig \pmatrix{ 1 \cr \ph} \; , 
\eeq
and the invariant is calculated as
\beq
  X = |\chi|^2 = |\sig|^2 ( 1 + |\ph|^2) \;.
\eeq
This provides the (canonical) line bundle over 
${\bf C}P^N$~\cite{Ca,PP,HKN4,HKN5}.

\item
$\psi \in {\bf N}$.\\
The standard representation is given by
\beq
 \chi = \rho (\xi) \pmatrix{ 0 \cr \psi } 
      = \xi \pmatrix{ 0 \cr \psi } 
      = \pmatrix{ 0 \cr \psi} \;.
\eeq
Hence the invariant $X = |\psi|^2$ is trivial, and 
this gives just the direct product of ${\bf C}P^{N}$ 
and the space of $\psi$.

\item 
$\psi \in \overline{{\bf N}}$.\\
The standard representation for 
the matter field $\psi$ is given by
\beq
 \chi = \rho(\xi) \pmatrix{ 0 \cr \psi}  
 = \xi^{-1 T} \pmatrix{ 0 \cr \psi} 
 = \pmatrix{ - \ph \cdot \psi \cr \psi } \;.
\eeq
Hence the invariant is calculated as
\beq
 X = |\psi|^2 + |\ph\cdot\psi|^2 \,.
 \label{CPN-X}
\eeq
The total space is (topologically) 
the ${\bf C}^{N}$-bundle over ${\bf C}P^{N}$ or 
the cotangent bundle over ${\bf C}P^{N}$, 
$T^* {\bf C}P^{N}$.\footnote{
This cotangent bundle does not have to be endowed 
with the hyper-K\"ahler metric.
}
We discuss this case in detail, below.

\end{enumerate}

After making some comments on the singlet matter $\sig \in {\bf 1}$ 
in the rest of this section, 
we will work out the anti-fundamental representation matter
$\psi \in \overline{\bf N}$ in detail in the next section. 

Since the invariant for the singlet can be rewritten as
\beq
 X = |\sig|^2 e^{K_0}  \;,\hs{5} 
 K_0 \equiv \log(1+|\ph|^2) \;,
\eeq
with $K_0$ the K\"ahler potential for the base ${\bf C}P^N$, 
the $G$-transformation law for $\sig$ is
\beq
 \sig \to \sig' = \sig e^{-\Lambda(g,\ph)} \;, \hs{5}
 K_0 \to K_0' = K_0 + \Lambda(g,\ph) + \Lambda^*(g,\ph^*)
\eeq
which is the one of the line bundle~\cite{line}. 
(Using the projection $\eta$, $\Lambda$ can be given by 
$\Lambda =  - \log \det_{\eta} \hat h'{}^{-1}$.)

If we define the function 
$g(X) \equiv f(X) + c \log X$, 
the \kahler potential for the total space can be rewritten as 
\beq
 K = c \log (1+|\ph|^2) + g(X) - c \log X 
   = g(X) - \log \sig - \log \sig^* \;,
\eeq 
where the last two terms can be eliminated by 
the \kahler transformation.
Therefore we do not need to include 
the \kahler potential 
for the base space ${\bf C}P^N$ itself
into the one for the total space. 
Then the Ricci-flat metric and its \kahler potential 
was obtained in \cite{HKN4,HKN5}. 
We just quote the result here:
\beq
 K = g (X) =  
\big( \lambda X^{N+1} + b \big)^{\frac{1}{N+1}} 
+ b^{\frac{1}{N+1}} 
\cdot I \big( b^{- \frac{1}{N+1}} 
\big( \lambda X^{N+1} + b \big)^{\frac{1}{N+1}} ; N+1 \big) \; , 
\label{kahler-Calabi}
\eeq
with $b$ an integration constant, $\lam$ a constant 
related to $N$, and $I(y;n)$ the function defined by
\beq
&&\hs{-13} 
 I (y; n) 
  \equiv  
  \int^{y} \! \frac{dt}{t^n - 1} \non
&& =  \frac{1}{n} \Big[ \log \big( y - 1 \big) 
    - \frac{1 + (-1)^n}{2} 
    \log \big( y + 1 \big) \Big] \non
&&
 + \frac{1}{n} \sum_{r=1}^{[\frac{n-1}{2}]} \cos \frac{2 r \pi}{n} 
\cdot \log \Big( y{}^2 - 2 y \cos \frac{2 r \pi}{n} + 1 \Big) \non
&&  + \frac{2}{n} \sum_{r=1}^{[\frac{n-1}{2}]} 
\sin \frac{2 r \pi}{n} 
\cdot \arctan \left[ \frac{\cos (2 r \pi / n) - y}
                    {\sin (2 r \pi /n) } \right] \; . \label{fn-I}
\eeq
We note that 
the coordinate transformation $\rho = \sig^{N+1}$, 
with $\ph$ unchanged, 
is needed to get a regular metric at $\sig =0$~\cite{HKN4}. 
Then the invariant in the new coordinates 
is $X^{N+1} = |\rho|^2 e^{(N+1) K_0}$. 
We thus find that this is the canonical line bundle.
In the limit of $b \to 0$, 
the manifold becomes an orbifold 
${\bf C}^{N+1}/{\bf Z}_{N+1}$~\cite{Ca,PP,HKN4,HKN5}. 
Therefore $b$ regularizes the orbifold singularity.

\medskip
\section{New Calabi-Yau metric on $T^*{\bf C}P^N$}
We work out the anti-fundamental representation 
$\psi \in \overline{\bf N}$ in this section. 
The manifold is topologically $T^*{\bf C}P^N$ 
but will not be equipped with a hyper-K\"ahler metric 
except for $N=1$. The relation with 
the hyper-K\"ahler metric on $T^*{\bf C}P^N$ 
will be discussed in Appendix. 
We denote coordinates of the total space by 
$z^{\mu} = (\ph^i, \psi_{\ib})$ ($i,\ib = 1,\cdots N$).
We represent the differentiations with respect 
to these coordinates by a comma with 
indices of corresponding coordinates. 
The differentiations of $f$ can be calculated, to give
\beq
 f,_i = f' \cdot \psi_{\ib} (\ph^* \cdot \psi^*) \,,\hs{5} 
 f,_{\ib} = f' \cdot [\psi^*_{\ib} + \ph^i (\ph^* \cdot \psi^*)]
\eeq
and 
\beq
 && f,_{ij^*} = f' \cdot \psi_{\ib} \psi^*_{\jb} 
 + f'' \cdot \psi_{\ib} \psi^*_{\jb} |\ph \cdot \psi|^2  \,,\non
 && f,_{i\jb^*} = f' \cdot \psi_{\ib} \ph^{*j} 
 + f'' \cdot \psi_{\ib} (\ph^*\cdot\psi^*) 
    [ \psi_{\jb} + \ph^{*j} (\ph \cdot \psi) ]  \,,\non
 && f,_{\ib\jb^*} =  f' \cdot (\delta_{ij} + \ph^i \ph^{*j})
 + f'' \cdot [ \psi^*_{\ib} + \ph^i (\ph^* \cdot \psi^*) ] 
       [ \psi_{\jb} + \ph^{*j} (\ph \cdot \psi) ] \,, \;\; 
   \label{dif-f}
\eeq
where the prime denotes the differentiation with respect to 
the argument $X$. 
Using these expressions, components of the metric 
$g_{\mu\nu^*} = K,_{\mu\nu^*}$ can be written as
\beq
 g_{\mu\nu^*} 
 = \pmatrix{g_{ij^*}    & g_{i\jb^*} \cr
            g_{\ib j^*} & g_{\ib\jb^*} } \, ,
\eeq
with each block being 
\beq
 && g_{ij^*} 
 = c \left[ { \delta_{ij} \over 1+|\ph|^2} 
       - {\ph^{*i} \ph^j \over (1+|\ph|^2)^2} \right]
  + f,_{ij^*}  \,,\non
 && g_{i\jb^*} =  c { \ph^{*i} \over 1+|\ph|^2} + f,_{i\jb^*} \,,\non
 && g_{\ib\jb^*} = f,_{\ib\jb^*} \;.  \label{metric}
\eeq

Let us calculate the determinant of this metric. 
As vacuum expectation values, 
we can set $\left<\ph \right> = 0$ 
and $\left<\psi \right> = (\epsilon , 0,\cdots,0)^T$ 
without loss of generality.\footnote{
Any point on the manifold can be transformed onto 
this point $v$ by a $G$-transformation; 
We can take $\left<\ph \right> = 0$, because $\ph$ parameterize 
homogeneous manifold $G/H$ and hence any $\ph$ 
can be transformed to the origin by a $G$-transformation. 
Then, we can set $\left<\psi\right>$ as one component 
using an $H$-transformation, because $\psi$ belongs 
to an irreducible representation oh $H$. 
} 
Then the determinant of the metric on this point $v$ is 
\beq
 \det g_{\mu\nu^*}|_v 
 = c^{N} (f')^{N-1} (c + f' |\epsilon|^2) (f' + f''|\epsilon|^2) \, .
\eeq
Using $X|_v = |\epsilon|^2$ at $v$, we obtain 
\beq
 \det g_{\mu\nu^*} 
 = c^{N} (f')^{N-1} (c + f' X) (f' + f''X) \, 
\eeq
for the determinant of the metric at general points 
because $X$ is the $G$-invariant.

Using the determinant of the metric, 
the Ricci-form can be written as  
$R_{\mu\nu^*} \equiv 
- \del_{\mu}\del_{\nu^*} 
\log \det g_{\kappa\lambda^*}$. 
The Ricci-flat condition $R_{\mu\nu^*}=0$ is 
in general a partial differential equation 
which is very difficult to solve. 
However, in this case, we can reduce it to ODE 
\beq
 (f')^{N-1} (c + f' X) (f' + f''X) = {a \over N+1}  \;, 
\eeq
with $a$ a real constant, 
and the numerical factor is just for convenience.

We can solve this ODE easily. 
Setting 
\beq 
F(X) \equiv f'(X)  X
\eeq
we obtain 
\beq
 F^{N-1} (c + F) F' = {a \over N+1}  X^{1-N} \;. 
\eeq 
This can be integrated, to give an algebraic equation
\beq
 F^{N+1} + {N+1 \over N} c F^{N} 
 = {a \over 2-N} X^{2-N} + b \;  \label{solution}
\eeq
for $N \neq 2$, or 
\beq
 F^3 + {3 \over 2} c F^2 = a \log X + b \label{solutionN=3}
\eeq
for $N=2$, 
with $b$ being a real integration constant. 
In principle Eq.~(\ref{solution}) can be solved numerically, 
but for lower $N$ we can solve it analytically. 

\medskip
First let us consider the $N=1$ case of $T^*{\bf C}P^1$. 
The invariant is $X= |\psi|^2 (1+|\ph|^2)$ for $N=1$, 
with $\ph$ and $\psi$ one component. 
Eq.~(\ref{solution}) reduces to 
\beq
 F^2 + 2 c F = a X + b \;,
\eeq
which can be solved
\beq
 F(X) = f' X = - c \pm \sqrt{c^2 + b + a X} \;.
 \label{F-sol1}
\eeq
This can be integrated once again, to yield
\beq
 f(X) = - c \log X 
  \pm \left[ 2 \sqrt{a X + r^2} 
  + r \log \left( \sqrt{a X + r^2 } - r 
            \over \sqrt{a X + r^2 } + r \right) \right] 
  \label{EH1} 
\eeq
with $r \equiv \sqrt{c^2 + b}$, 
where we have not included an integration constant 
because it does not contribute to the metric.
Therefore we obtain 
\beq
 K &=& c\log (1+|\ph|^2) + f(X) \non
   &=& 2 \sqrt{a X + r^2} 
  + r \log \left( \sqrt{a X + r^2 } - r 
            \over \sqrt{a X + r^2 } + r \right) 
  - c \log \psi - c \log \psi^* \;, \label{EH2}
\eeq
where the last two terms can be eliminated by 
the K\"ahler transformation.
Here we have chosen the plus sign in Eq.~(\ref{EH1}) for 
the positivity of the metric. 
Setting $\ph' = \psi \ph$ 
the invariant is $X=|\psi|^2 + |\ph'|^2$.
Eq.~(\ref{EH2}) is the K\"ahler potential \cite{GP} for 
the Eguchi-Hanson metric \cite{EH} 
on $T^*{\bf C}P^1$. 
This is a hyper-K\"ahler manifold 
because any Ricci-flat \kahler manifold is 
a hyper-K\"ahler manifold in real four-dimensions
[the holonomy is $SU(2) \simeq Sp(1)$]
but it is not true for higher $N$.

\medskip
Next let us consider the $N=2$ case of $T^* {\bf C}P^2$.
For $N=2$, third order Eq.~(\ref{solutionN=3}) 
can be solved, to give 
\beq
 F(X)  = \cases{ 
 - { c \over 2} + G_+^{1\over 3}(X) + G_-^{1\over 3}(X) \;, \cr 
 - { c \over 2} + \omega^2 G_+^{1\over 3}(X) 
                + \omega^3 G_-^{1\over 3}(X) \;, \cr 
 - { c \over 2} + \omega^3 G_+^{1\over 3}(X) 
                + \omega^2 G_-^{1\over 3}(X) \;, 
} \label{F-sol2}
\eeq
with $\omega = e^{2 \pi i /3 }$ and 
\beq
 G_{\pm}(X) \equiv \1{2}\left[
  a \log X + b - {c^3 \over 4} 
 \pm \sqrt{ (a \log X + b)^2 - {c^3 \over 2}(a \log X + b)}\,
\right] \;. 
\eeq
When the discriminant is positive, 
the first one is the real solution. 
Otherwise the rest ones are needed. 
We thus obtain (for the positive discriminant)
\beq
 K = c \log(1+|\ph|^2) - {c \over 2} \log X 
   + \int dX  X^{-1} (G_+^{1\over 3} + G_-^{1\over 3} ) \;.
\eeq
This is no longer hyper-K\"ahler. 
We have obtained a Calabi-Yau (but not hyper-K\"ahler) metric 
on $T^*{\bf C}P^2$.

\medskip
In the end, we consider the
$N=3$ case of $T^* {\bf C}P^3$. 
Eq.~(\ref{solution}) reduces to 
\beq
 F^4 + 4 c' F^3 + a X^{-1} - b = 0
\eeq
with $c' = {1 \over 3}c$. 
Then $F$ has four solutions given by
\beq
 && F(X) = \cases{ - c' \mp \1{2} \sqrt{\lam - p} 
  + \1{2} \sqrt{ -\lam - p \pm 2q (\lam -p)^{-\1{2}}} \;, \cr
      - c' \mp \1{2} \sqrt{\lam - p} 
  - \1{2} \sqrt{ -\lam - p \pm 2q (\lam -p)^{-\1{2}}} \;, 
  }  \label{F-sol3}
\eeq
with
\beq
 && 
  \lam = - 2 {c'}^2 
  + \1{3} \left[\1{2} \left(-q + \sqrt{q^2 + 4p^3}\right)\right]^{\1{3}}
  + \1{3} \left[\1{2} \left(-q - \sqrt{q^2 + 4p^3}\right)\right]^{\1{3}} 
   \;,\\
 && 
  p = -{4\over 3} (aX^{-1} + b ) \;,\\
 && 
  q = - 16 ( a c'{}^2 X^{-1} + 2 b c'{}^2 - 8 c'{}^6 ) \;.
\eeq
The K\"ahler potential for the total space is obtained
\beq
 K = c \log (1+|\ph|^2) - {c \over 3} \log X 
   + \int dX X^{-1} (\cdots) \;,
\eeq
where dots denote the last two terms 
in $F$ in the solution (\ref{F-sol3}).

To write down the metric, we do not need 
the K\"ahler potential itself 
but its derivative $f'$. 
We thus have derived the explicit expression for the metric 
for $N= 2,3,4$.

\section{Summary and Discussions}
We have given a method to construct 
a $G$-invariant metric on the vector bundle over K\"ahler $G/H$ 
using matter coupling of the nonlinear realization. 
The dimension of the vector as fiber is restricted 
by the $G$-invariance of the manifold because  
the vector belongs to the $H$-representation. 
To solve the Ricci-flat condition, 
cohomogeneity one is essential 
in which case it is reduced to 
ODE to be solved easily.
This requirement implies that  
the matter should belong to an irreducible 
representation of $H$ at least. 
As a concrete example, 
we have discussed the matter coupling in the ${\bf C}P^N$ model. 
The singlet matter has reproduced 
the canonical line bundle and
the anti-fundamental representation has given 
a Calabi-Yau metric on $T^* {\bf C}P^N$ 
which is not hyper-K\"ahler.
We have given the explicit expression for 
the metric for $N= 2,3,4$ [Eq.~(\ref{metric}) 
with (\ref{F-sol1}), (\ref{F-sol2}) and (\ref{F-sol3})] and 
the algebraic equation (\ref{solution}) for higher $N$.
A relation with hyper-K\"ahler Calabi metric on $T^* {\bf C}P^N$ 
is given in Appendix A.

We could introduce the matter belonging to 
higher (irreducible) representation of $H$ 
which would provide a new metric. 
Our model can be extended to other base manifolds 
for instance to the quadric surface 
$Q^N=SO(N+2)/[SO(N) \times U(1)]$~\cite{HN1,HN3,HKNT}. 
We will reproduce the canonical line bundle over $Q^N$ 
and will get the Calabi-Yau metric on $T^* Q^N$ 
which is not hyper-K\"ahler.

Our model is a finite nonlinear sigma model 
because the beta function is proportional 
to the Ricci-form.
For the ${\bf C}P^N$ model, 
$R_{\mu\nu^*} = h g_{\mu\nu^*}$ with positive $h$ holds
and so it is not finite. 
Instead, there exists a mass gap and 
a gauge boson is dynamically induced 
in the large-$N$ limit~\cite{CPN}.
We added matter into the ${\bf C}P^N$ model 
to make total finite. 
Investigating the relation with quantum properties of 
the ${\bf C}P^N$ model 
and our model is interesting. 
Looking for a conformal field theory 
corresponding to our finite model is also an important task.

\section*{Acknowledgements} 

The author thanks Kiyoshi Higashijima and Tetsuji Kimura 
for a collaboration in early stages of this work. 
He is also grateful to Thomas E. Clark for 
some comments on finiteness and 
a relation with the ${\bf C}P^N$ model.
His work is supported by the U.~S. Department 
of Energy under grant DE-FG02-91ER40681 (Task B).

\begin{appendix}


\section{Relation with the hyper-K\"ahler 
Calabi metric 
}
The standard metric on $T^* {\bf C}P^N$ is 
the Calabi metric~\cite{Ca}, 
which is a hyper-K\"ahler metric.
Here, we discuss the relation between our metric and 
the Calabi metric using 
a (hyper-)K\"ahler quotient construction~\cite{HKQ} 
for $T^* {\bf C}P^N$~\cite{HKQ-CPN,ANNS}. 
First prepare chiral superfields $\phi$ and $\chi$ 
belonging to ${\bf N+1}$ and $\overline{\bf N+1}$ of $SU(N+1)$,
respectively. 
Let $V$ and $\sigma$ be auxiliary vector and chiral 
superfields, respectively, 
introduced as Lagrange multipliers to give constraints 
among $\phi$ and $\chi$. 
Consider a $U(1)$ gauge symmetry, given by 
\beq 
 && V \to V' = V - i \Lambda + i \Lambda\dagg \;, \hs{5}
    \sig \to \sig' = e^{-i (1+q) } \sig \;, \non
 && 
 \phi \to \phi' = e^{i \Lambda} \phi \;, \hs{5}
 \chi \to \chi' = e^{i q \Lambda} \chi \;,
\eeq
where $q$ is the $U(1)$-charge for $\chi$ relative to $\phi$.
Then, the most general Lagrangian for these field contents 
can be written as~\footnote{
At first sight, one might consider that 
the most general K\"ahler potential should be
${\cal F}(e^V \phi\dagg\phi,e^{q V}\chi\dagg\chi)$.
However we can show that one variable in 
the arbitrary function can be linearlized 
in the path integral formalism~\cite{HN2}.
}
\beq
 {\cal L} 
 = \int d^4 \theta 
  \left[e^V \phi\dagg\phi + f (e^{q V}\chi\dagg\chi) - c V \right]
 + \left[\int d^2 \theta\; \sigma (\phi \cdot \chi - \delta_{q,-1} b) 
          +{\rm c.c.} \right]\;,
\eeq
with $f$ an arbitrary function. 
Here $c \in {\bf R}$ is a Fayet-Iliopoulos parameter and
$b \in {\bf C}$ can be non-zero only when $q = -1$.

We discuss particular values of $q$: 1) $q=0$ and 2) $q=-1$.\\
1) The \kahler potential on $T^*{\bf C}P^N$ given by 
${\bf C}P^N$ coupled with 
the matter belonging to $\overline{\bf N}$ 
corresponds to the case of $q=0$ (and $b=0$):
\beq
 {\cal L}_{q=0} = \int d^4 \theta 
  \left[e^V \phi\dagg\phi + f (\chi\dagg\chi) - c V \right]
 + \left[\int d^2 \theta\; \sigma (\phi \cdot \chi) +{\rm c.c.} \right]\;.
\eeq
The equations of motion for $V$ and $\sigma$ read
\beq
 {\del{\cal L} \over \del V} = e^V \phi\dagg\phi  - c = 0 \;, \hs{5}
 {\del{\cal L} \over \del \sigma} = \phi \cdot \chi = 0 \;,
\eeq
respectively.
Eliminating $V$ and solving the constraint, we obtain
\beq
 {\cal L}_{q=0} 
 = \int d^4 \theta 
  \left[ c \log \phi\dagg\phi + f(\chi\dagg\chi) \right] \;,
\eeq
with a gauge fixing $\phi^1=1$:
\beq
 \phi = \pmatrix{1 \cr \ph} \; \hs{5} 
 \chi = \pmatrix{ - \ph \cdot \psi \cr \psi} \;. \label{fixing}
\eeq
This is the \kahler potential considered in the above discussion, 
Eq.~(\ref{CPN-matter}) with (\ref{CPN-X}).

~\\
2) On the other hand, the hyper-K\"ahler metric on 
$T^*{\bf C}P^N$ can be obtained by choosing $q=-1$:
\beq
 {\cal L}_{q=-1} 
 = \int d^4 \theta 
  \left[e^V \phi\dagg\phi + f (e^{-V}\chi\dagg\chi) - c V \right]
 + \left[\int d^2 \theta\; \sigma (\phi \cdot \chi-b) 
         + {\rm c.c.} \right]\;.
\eeq
We should impose the Ricci-flat condition to determine $f$ 
after eliminating $V$, but we know the answer $f(X)=X$ because 
it is hyper-K\"ahler: 
\beq
 {\cal L}_{q=-1,RF} 
 = \int d^4 \theta 
  \left[e^V \phi\dagg\phi + e^{-V}\chi\dagg\chi - c V \right]
 + \left[\int d^2 \theta\; \sigma (\phi \cdot \chi-b)  
         + {\rm c.c.} \right]\;.
\eeq
This is the hyper-K\"ahler quotient construction~\cite{HKQ} 
for the hyper-K\"ahler Calabi metric~\cite{Ca} 
on $T^* {\bf C}P^N$~\cite{HKQ-CPN,ANNS}.\footnote{
Adding masses to hypermultiplets, the potential term 
for $T^*{\bf C}P^N$ can be obtained~\cite{ANNS}.
} 
The ${\cal N}=2$ SUSY is enhanced to 
${\cal N}=4$ SUSY in two-dimensional space-time 
(${\cal N}=2$ SUSY in four-dimensional space-time) 
with $(V,\sigma)$ ${\cal N}=4$ vector multiplet 
and $(\phi,\chi)$ ${\cal N}=4$ hypermultiplets.
$c$ and $b$ constitute the triplet of the FI parameters.
After eliminating $V$ and $\sigma$ by their equations 
of motion, we obtain
\beq
 K = c \log |\phi|^2 
   + \sqrt{c^2 + 4 |\phi|^2 |\chi|^2 } 
   - c \log \left(c + \sqrt{c^2 + 4 |\phi|^2 |\chi|^2 } \right)
   \;,
\eeq
with proper gauge fixing. (For $b=0$ we have (\ref{fixing}), 
and for $b \neq 0$ we should take 
an another gauge~\cite{HKQ-CPN,ANNS}.)

We thus have found the difference between 
our Calabi-Yau metric 
and the hyper-K\"ahler Calabi metric on $T^* {\bf C}P^N$ 
comes from the relative gauge $U(1)$-charge 
between $\chi$ and $\phi$.

For the general value of $q$, 
the \kahler potential after eliminating $V$ is
\beq
 K = c \log \phi\dagg\phi 
   + h[(\phi\dagg\phi)^{-q} \chi\dagg\chi] \;,
\eeq
with some function $h$ related with $f$. 
This can be found from 
the algebraic geometry point of view~\cite{LT}: 
the argument $(\phi\dagg\phi)^{-q} \chi\dagg\chi$ of $h$ 
is the gauge invariant and remains 
after the integration over $V$.
The construction of the Calabi-Yau metric with 
general $q$ is left for a future work.

\end{appendix}


\end{document}